\documentstyle[preprint,aps]{revtex}
\tightenlines
\begin{document}
\draft
\title{Orthopositronium: ``on the possible relation of gravity to electricity''}
\author{B. A. Kotov$^{\dagger}$, B. M. Levin $^{1}$, V.I.Sokolov $^{2}$}
\address{A.F. Ioffe Physical-Technical Institute, 194021 St. Petersburg, Russia}
\maketitle

\begin{abstract}
The resolve of the `orthopositronium-lifetime puzzle' needs more study of
the ``isotope anomaly'' in gaseous neon and of the contribution $\sim $ 2$%
\cdot 10^{-3}\,$of non-perturbative mode into orthopositronium annihilation.

The Michigan results (2003) are considered as {\it the first supervision of
relation between gravitation and electricity.}

For the decision of alternative in interpretation of new and former results
it is necessary to execute the program of additional measurements.
\end{abstract}


\section{THE ORTHOPOSITRONIUM-LIFETIME PUZZLE IS NOT SOLVED: ON AN EFFECT OF
NON-PERTURBATIVE CONTRIBUTION}

As it known [1], a workgroup of Michigan University (Ann Arbor, USA)
published their new experimental results of vacuum measurements of
orthopositronium ({\it o}-Ps, $^{\text{T}}$Ps) annihilation decay rate $%
\lambda _{\text{T}}$\thinspace \thinspace (lifetime, decay rate 1/$\lambda _{%
\text{T}}$). They headed this article `Resolution of
Orthopositronium-Lifetime Puzzle'.

\noindent `{\it Our new value for the o}-Ps {\it is then} $\lambda _{\text{T}%
}$ = 7.0404(10)(8) $\mu $s$^{-1}$, {\it where the first error is the
statistical error of }140 ppm {\it and the second error} {\it of} 115 ppm
{\it represents a combined systematic error }%
\mbox{$<$}
. . .%
\mbox{$>$}
{\it our current value for }$\lambda _{\text{T}}$ {\it using a porous silica
thin film for production of near-thermal }Ps is {\it in excellent agreement
with theory at a combined level of precision of} 180 ppm. {\it It also
agrees well with the Tokyo }[2] {\it experiment}', they wrote in the article
[1].

This `excellent agreement with theory' provides calculating the ground state
of electron-positron atom according to the perturbative calculations in
Quantum Electrodynamics (QED, perturbative dynamics). At the present, these
calculations achieved to take into account total corrections of the order $%
\alpha ^2$ and a precision of 1.6$\cdot 10^{-4}\%$\thinspace \thinspace [3].

From the beginning of 1980's and until the middle of 1990's, this Michigan
group arrives to a high graduate of precision to measure annihilation rates
of the ({\it o}-Ps $\lambda _{\text{T}}$) and of the parapositronium ({\it p}%
-Ps, $^{\text{S}}$Ps; $\lambda _{\text{T}}$), go into details with [4-7]. It
should be noted that they used unique experimental technique, which resulted
the $\lambda _{\text{T}}$ precision of 0.02\% [5] in buffer gases
experiments and the precision of 0.023\% using a vacuum technique. As a
result they established a discrepancy between theory and experiment that was
the experimental $\lambda _{\text{T}}$\thinspace \thinspace exceeding on $%
\triangle \lambda _{\text{T}}$\thinspace \thinspace its calculated value
with 9.4$\sigma $\thinspace \thinspace in the gaseous experiments [5] and
with 6.2$\sigma $\thinspace \thinspace in the vacuum experiments [6].These
experimental facts stimulated so much experimental statements and the QED
large corrections calculation's during the latest two decades, so those
stimulated new approaches to solve the `orthopositronium-lifetime puzzle'.

The authors of [1] think that sources of previous experiments' systematic
measurement errors are founded and these have a real base [8, 9]. So the new
measurement result finishes the problem to deviate the o-Ps experimental
annihilation rate with theory, they think. In this way, the previous results
of Michigan group should be refused, so the all two decades' attempts to
solve the puzzle should be regarded without ground. The Tokyo group's
experimental result [2] also supported this conclusion.

However basing on facts that `isotope anomaly' {\it o}-Ps [10] and a
non-perturbative contribution $\triangle \lambda _{\text{T}}/\lambda _{\text{%
T}}\cong 0.19\%$\thinspace \thinspace into {\it o}-Ps dynamic (where {\it o}%
-Ps is derived from $\beta ^{+}$-decay positrons [11-14] in the final state
of the topological quantum transition) are exist, our opinion lead us to a
critical analysis of the new measurements [1]. So the next our thesis needs
to ground:

The rate of o-Ps self-annihilation $\lambda _{\text{T}}=7.0514(14)\mu $s$%
^{-1}$\thinspace \thinspace was measured in the buffer gases experiments [5]
and $\lambda _{\text{T}}=7.0482(16)\mu $s$^{-1}$\thinspace \thinspace in
vacuum experiment [6], received earlier by Michigan group conserves real
status.

The start point of the critical analysis is the statement contained in [1]:

`{\it The Michigan buffer gas experiment} [5] {\it has been shown to be
subject to the problem of incomplete} Ps {\it thermalization} [8] {\it in
low-pressure gases, and this value should be corrected downward by at least}
700 ppm'.

To the contrary we have another opinion: the experimental method in the
buffer gases [5] gave a possibility to measure the best value of the rate of
{\it o}-Ps self-annihilation, because this measurements in magnetic field
produced complete experience that is measurements of {\it o}-Ps and {\it p}%
-Ps self-annihilation rates at the same time, so the new data [1] do not
change this situation because the data [1] were received in qualitative
another experiment.

\noindent The point is that, from viewpoint of the contribution a
non-perturbative dynamics in {\it o}-Ps, formed by $\beta ^{+}$-decay
positrons ($^{22}$Na, $^{68}$Ga) [11-14], 64\% {\it o-}Ps sub-state (m=0),
annihilated into 2$\gamma $-quantum (at {\it B}=2.85 kG [8,9]), are
essentially different in the thermalization from 36\% o-Ps (m=0) sub-state
and {\it o}-Ps (m=$\pm 1$) sub-state, testing three-quantum annihilation.
These latest have not even a term `kinetic energy', because an additional
factor of the `mirror Universe' is included owing to presence at their
dynamics the single-photon (virtual) annihilation. Or else, `kinetic energy'
of the {\it o}-Ps (m=$\pm $1) sub-states and 36\% {\it o}-Ps (m=0) sub-state
are compensated by an exchange {\it o}-Ps$\Leftrightarrow ${\it o}-Ps$%
^{^{\prime }}$\thinspace \thinspace with the `mirror Universe', but the {\it %
o}-Ps (m=0) sub-state (2$\gamma $-annihilation) in an exchange with the
`mirror Universe' does not participate [11, 12]. Therefore in the
Doppler-broadening measurements of the 511 keV-annihilation line (2$\gamma $%
-annihilation) [8,9] it is possible to receive only information about
thermalization of the short-lived share of the {\it o}-Ps (m=0) sub-state,
but not about its long-lived share and {\it o}-Ps (m=$\pm $1) sub-state.
Thus, statement of the authors [1] that in work [8] regular mistake of their
former measurement is experimentally proved [5] ({\it i.e.} excess of
measured $\lambda _{\text{T}}$ size, `at least 700 ppm'), can be denied from
positions of the account of a non-perturbative mode. Really, {\it o}-Ps
magnetic momentum is zero, and {\it p}-Ps magnetic momentum equals two Bohr
magnetons, {\it o}-Ps energy exceeds {\it p}-Ps energy on size
$$
\triangle \text{W}=\text{W}\alpha ^2\cdot (\frac 43+1)\cong 8.4\cdot 10^{-4}%
\text{eV}
$$

\noindent (hyperfine splitting of the Ps ground states; W$\cong 6.8$%
\thinspace \thinspace eV-- binding energy of the Ps), so at {\it o}-Ps birth
(and collapse-reduction of its wave function at interactions in a gas) a
very tiny energetic shift occurs

$$
\begin{array}{c}
^{
\text{3}}\text{{\it E}(m=0)--}^{\text{3}}\text{{\it E}(m=}\pm \text{1)=}%
\frac{\triangle \text{W}}2\cdot \left[ (1+x^2)^{1/2}-1\right] \cong \\ \cong
10^{-6}\text{eV ({\it B=}2.85 kG),}
\end{array}
$$

\noindent where
$$
x=\frac{2e\hbar }{m_e}\cdot \frac B{\triangle \text{W}}=2.75\cdot
10^{-2}B\,\,\text{({\it B }in{\it \ }kG).}
$$

\noindent Because {\it o}-Ps (m=$\pm $1) does not ``sense'' magnetic field
and the 64\% of {\it o}-Ps (m=0) under {\it B}=2.85 kG does not `sense' the
``mirror Universe'', so energetic relaxation between {\it o}-Ps sub-states
is possible as a result of solely collisions in gas. However this process is
difficult because of kinematics (see, for example, [15]). So, the
Doppler-broadening measurements of the 511 keV-annihilation line do not
produce information on the {\it o}-Ps thermalization. At the present we have
not a method to measure a kinetic energy (thermalization) of the {\it o}-Ps
(three-quantum annihilation, wide distribution of energy -- from E$_\gamma
\sim 0$ up to E$_\gamma \sim 2m_ec^2$) produced by $\beta ^{+}$-decay
positrons in substance.

Actually, from standard QED viewpoint (perturbative dynamics) the supposed
deviation of thermalization rates of {\it o}-Ps (m=0) sub-state (2$\gamma $
-annihilation) and {\it o}-Ps (m=$\pm $1) sub-state (3$\gamma $
-annihilation) is absent. However we are going to consider the experimental
results [1], so any a priori considerations even ``evidently''
considerations shall be excluded.

As it was mentioned above, it was a principle methodical feat that the
studies of 1982-89 [4,5] had got a non-beaten measurement precision and had
excluded systematic errors to measure parameters of {\it o}-Ps lifetime
spectra annihilation under moderate magnetic fields of {\it B} = 3.5 kG. The
main goal of this method is that {\it o}-Ps and {\it p}-Ps annihilation
rates were measured in the same experiment. To realize full experiment
implies to exclude an affect, which an environment does on wave function of
Ps (the affect shall be excluded with high degree of precision). As it is
evidently, such affect is unavoidable in substance (polarization forces).
Tokyo group did not take into account this principle factor in their works.
An observable rate of {\it o}-Ps annihilation decay in substance is defined
in [2] as follows
$$
\lambda _{obs}(t)=\left| \Psi _{obs}(t)\right| ^2/\left| \Psi (0)\right|
^2\cdot \lambda _{\text{T}}+\lambda _{pickoff}\,(t),
$$

\noindent where $\left| \Psi _{obs}(t)\right| ^2$ and $\left| \Psi
(0)\right| ^2$ are densities of the wave function in substance and vacuum,
accordingly, $\lambda _{pickoff}$ -- the {\it o}-Ps pickoff-annihilation
rate, and the both sign of the inequalities are possible $\left| \Psi
_{obs}(t)\right| ^2/\left| \Psi (0)\right| ^2$%
\mbox{$<$}
1 or $\left| \Psi _{obs}(t)\right| ^2/\left| \Psi (0)\right| ^2$%
\mbox{$>$}
1 (Ps is ``swelling'' or ``compressed'' by polarization forces in substance,
see [15]). Therefore used in work of Tokyo group approximation [2] $\left|
\Psi _{obs}(t)\right| ^2/\left| \Psi (0)\right| ^2$=1 to describe {\it o}-Ps
annihilation in an aerogel powder had not ground, a systematic error these
measurements made is much underestimated most likely (indirectly it is
marked also in [1]). A physical mechanism of this distortion is an affect of
electric field that is Stark effect. This displays itself in {\it o}-Ps
dynamics specific way: electric field selects some specific direction
(linear polarization of {\it o}-Ps) that excludes a possibility to realize
supersymmetric mode ($\sim $0.2\%) and hence annihilation with notoph ($%
\gamma ^{\circ }$) o-Ps%
\mbox{$\backslash$}
o-Ps$^{\prime }$(p-Ps$^{^{\prime }}$)$\,\rightarrow \,\gamma ^{\circ
}\backslash 2\gamma ^{^{\prime }}$ [11,12] is excluded. If {\it o}-Ps is
localized in pore of aerogel, then electric field (which the symmetry
breaks) acts on the {\it o}-Ps wave function in a double electrical layer on
the boundary solid phase - vacuum in the pore. New more exact comparative
measurements of the Tokyo group with use hydrophilic and hydrophobic samples
of aerogel SiO$_2$ [16] do not change a conclusion about suppression in
these conditions of the contribution of non-perturbative (supersymmetric)
mode.

Being ones developed the method [1], they targeted to arrive to a condition
of {\it o}-Ps thermalization. Having this task, they used an electric field $%
\sim 4\cdot 10^3$ V/cm to deviate a positronium bunch 700 eV (produced by $%
\beta ^{+}$-decay of $^{22}$Na) onto a porous target along 0.75 cm. The
consideration to crush {\it additional supersymmetric mode} {\it o}-Ps%
\mbox{$\backslash$}
{\it o}-Ps$^{\prime }$({\it p}- Ps$^{\prime }$) with electric field we
mentioned above give a possibility (viewpoint of non-perturbative dynamics)
to prefer the Michigan group first experiment [6], because this first
experiment had not contained an affect of electric field to {\it o}-Ps
annihilation in contrast to the new method [1].

In general, taking into account that non-perturbative dynamics give a
contribution, the problem to thermalization {\it o}-Ps is absent, so the
first experiments Michigan group had made [4-6] give the rate of
self-annihilation of {\it o}-Ps, which includes an additional mode $\sim $%
~0.2\% (derived from $\beta ^{+}$-decay positrons in substance) completely
adequate.

A problem statement that annihilation characteristics {\it o}-Ps is linked
to a $\beta ^{+}$-decay nature of positrons was supposed in the second half
of 1970's and the beginning of 1980's. A ground of the supposition were
experimental observations of some peculiarities in lifetime spectra of
annihilation of positrons that were derived from $\beta ^{+}$-decay of $%
^{22} $Na ($^{22}$Na $\rightarrow \,\,^{22*}$Ne + e$^{+}$ + $\nu $) in {\it %
gaseous neon}. A ``start'' of the experiments (the moment of $\beta ^{+}$%
-decay final state: daughter nucleus $^{22}$Ne, positron (e$^{+}$), neutrino
($\nu $), formation Ps in substance) was detected using a nuclear $\gamma _{%
\text{n}}$-quantum ($^{22*}$Ne $\rightarrow ^{22}$Ne + $\gamma _{\text{n}}$%
). In the middle of 1980's the supposition was confirmed by observations of
``isotope anomaly'' of {\it o}-Ps [10]. Independent discovery of ``$\lambda
_{\text{T}}$ -anomaly'' and its confirmation [4-6] powered arguments in
behalf of non-perturbative dynamics contribution into {\it o}-Ps
annihilation, because the old calculations [17] to contribute additional
supersymmetric mode in {\it o}-Ps annihilation gave a possibility to
calculate a contribution of single-notoph annihilation contribution of
single-notoph annihilation {\it o}-Ps%
\mbox{$\backslash$}
{\it o}-Ps$^{\prime }$({\it p}-Ps$^{^{\prime }}$)$\,\rightarrow \,\gamma
^{\circ }\backslash 2\gamma ^{^{\prime }}$ (0.19\%) [11,12].

Taken into account the formed situation we considered above, the fact that
Michigan group refused [1] their previous experimental conclusions [4-6] can
be accepted as `resolution of orthopositronium-lifetime puzzle', so we can
formulate the next alternative:

1. If we be not left the frames of standard QED [3], then the new
experiments Michigan group made [1] result only confirmation of a systematic
error they had got in their experimental methods to measure the
selfannihilation rate [4-6].

\noindent And to the contrary:

2. To accept that ``isotope anomaly'' in neon [10] and the non-perturbative
dynamics' contribution (0.19[5]$\div $0.14[6])\% into {\it o}-Ps
annihilation [11,12] (where {\it o}-Ps is derived from $\beta ^{+}$-decay
positrons) have a general nature, lead to accept that unique original
methods and the obtained experimental results of ``$\lambda _{\text{T}}$%
-anomaly'' [4-6] are adequate to reality.

\noindent A way to solve ``orthopositronium puzzle'' (the orthopositronium
problem) as whole is more to study `isotope anomaly' {\it o}-Ps in neon [10]
and to set up experimentum crucis, a concept of which was formulated earlier
[11, 12] as follows:

To found a contribution of supersymmetric single-notoph mode of the
annihilation {\it o}-Ps%
\mbox{$\backslash$}
{\it o}-Ps$^{\prime }$({\it p}-Ps$^{^{\prime }}$)$\,$ needs a method
guaranteeing 4$\pi $-geometry to detect notoph.

A similar method was realized in [18] before. So, these are necessary:

$\bullet \,\,$to modify the method for positrons source $^{22}$Na in gaseous
neon of sufficiently high pressure, and also

$\bullet \,\,$to comparing measurements of neon tests of different isotope
mix. Our supposition based on observations of ``isotope anomaly'' of {\it o}%
-Ps (the contribution of non-perturbative dynamics) is that expected affect
shall be selected on a background of a pick total energy of {\it o}-Ps
annihilation under {\it E}$_{\gamma ^{\circ }}\cong 2m_ec^2\cong 1.022$ MeV,
not under {\it E}$_{\gamma ^{\circ }}\sim 0$ as the study [18] did it.

\section{MODEL OF FUNDAMENTAL SPACE-LIKE STRUCTURE WITH TWO-DIGIT PLANCK
MASS M$_{\mu }=\pm $M$_{\text{Pl}}$}

\medskip The citation in the preprint title repeats the appellation of the
paragraph in ``Experimental Researches in Electricity'' by M. Faraday where
negative results of the first experimental searches the relation of gravity
and electricity are stated [19]. These searches were stimulated by `the long
and constant
\mbox{$<$}
Faraday's%
\mbox{$>$}
persuasion that all the forces of nature are mutually dependent, having one
common origin' (Par. 2702). His comment full of self-respect, when the first
experimental attempts to establish prospective relation did not give
positive result, inspires today (Par. 2717): `{\it Here end my trials for
the present. The results are negative. They do not shake my strong feeling
of the existence of a relation between gravity and electricity, though they
give no proof that such a relation exists}'.

We examine a new precision measurement of the orthopositronium
selfannihilation decay rate $\lambda _{\text{T}}$ [1] as the first
supervision of this relation {\it between gravity and electricity}.

A feature of a new technique of measurement $\lambda _{\text{T}}$ in
comparison with the techniques realized earlier by Michigan group
(1982-1994) is the introduction in the measuring chamber of electric field.
This innovation and result received in the [1] allowing us to present an
orthopositronium problem in a context of a relation {\it between gravity and
electricity}. The base for such interpretation serves the phenomenology,
advanced for the description `isotope anomaly' and `$\lambda _{\text{T}}$%
-anomaly' {\it o}-Ps, which are examined as manifestation of long-range for
baryon charge [10-12]. Though the experiments made up to the middle of 90th,
have excluded a hypothesis about stationary long-range for baryon charge
(``fifth force''), it does not mention non-stationary aspect of similar
long-range in $\beta ^{+}$-decay, which is a source of the positrons forming
Ps, and is examined by us as {\it topological quantum transition} (TQT)
[10-12].

In this section the conception of a question are submitted. In following
section (III) the offer of the crucial experiment is presented.

\noindent 1. As it known [20] `...{\it in the equations of the modern theory
there is no equivalence of all speeds of movement as the whole (including
super-light speeds). As symmetry of the equations can be higher than
symmetry of the ground state, it is represented rather expedient to expand
symmetry of the equations up to a full relativity, i.e. equivalence of all
speeds (rule out the speed of light itself). In usual conditions to provide
sub-light character of relative speeds, \underline{additional symmetry}
should be broken spontaneously}' (it is underlined by us).

This theory has opened a prospect to an alternative (more precisely,
\underline{additional}) interpretation of faster-than-light speeds as a
space-like `{\it zero-mass Goldstone field which in linear approximation in
any way does not prove in classics and its unique macroscopical
manifestation is presence non-electromagnetic long-range interactions of
bodies with not disappearing average spin density.
\mbox{$<$}
...%
\mbox{$>$}
Restoration
\mbox{$<$}
the symmetry%
\mbox{$>$}
should be accompanied by doubling of space-time dimension}'.

In order to prevent the conflict with the general relativity (GR) additional
Goldstone field (GF) cannot cover ``all'' space-time; GF should be limited,
be peculiar ``defect'' of space-time. Besides, there is a problem of
causality. Both these problems are solved in the following point.

\noindent 2. `{\it Physical interpretation of some algebraic structures of
an energy-pulse tensor allows to assume, that the form of substance named }$%
\mu ${\it -vacuum, macroscopic possessing properties of vacuum is possible
\mbox{$<$}
...%
\mbox{$>$}
The homogeneous }$\mu ${\it -vacuum world has the de Sitter metrics}' [21].

The concept vacuum-like states of matter (VSM) is included in modern
cosmology. In [21] VSM are submitted as the homogeneous spherical de Sitter
worlds with positive curvature (K
\mbox{$>$}
0 -- de Sitter space of the 1st sort). Such world is `not empty' also a sign
on density of mass coincides with sign of K. Limitedness of the de Sitter
space the major quality which allows to use this GR mathematical device for
the description of space-like structures of TQT in final state of $\beta
^{+} $-decay. De Sitter 2nd sort spaces (K
\mbox{$<$}
0 -- with negative density of mass) in basic work [21] are excluded from
consideration as it seemed, that Riemann space of constant negative
curvature has properties physical interpretation of which is rather
difficult owing to infringement of physical causality (closed time-like
geodesic line). Meanwhile, published year earlier the theory of a discrete
scalar field with negative mass density (`C-field' [22]) allows, on the one
hand, to use it as a field compensating VSM in final state of TQT, and with
another -- to structure GF (cellular structure of VSM).

The analysis executed soon after publication [21], has shown that probably
`the simultaneous creation of quanta of fields of positive energy field and
of the negative energy {\it C}-fields' ([23], p. 90), because for a probe
particle `there are closed timelike lines in this space; however it is not
simple connected, and if one unwraps the circle {\it S}$^{\prime }$ (to
obtain its covering space {\it R}$^{\prime }$) one obtains the universal
covering space of anti-de Sitter space which does not contain any closed
timelike lines' ([23], p.131).

The opportunity of the development of the concept of a full relativity [20]
contains in a hypothesis about additional realization of supersymmetry. As
is known, double application of supertransformation -- from fermion to boson
and again to fermion -- translates a particle in other point of space. It
supposes additional treatment of mathematical structure of supersymmetry as
structures of long-range of the new type (non-Newtonian, non-Coulombian) in
final state of TQT [12]. All this concerns to a positron in {\it o}-Ps,
because electron is in so-called {\it entangled state} with all electrons
the observable Universe.

At last, also the problem of relativistic invariance (physical causality)
for examined GF (VSM) `as a whole' is solved naturally: non-locality in this
context does not result to a breaking of the causality, because VSM cannot
be a frame of reference (see [21]).

\noindent 3. It is necessary to prove a presence at cells (`units') of VSM
baryon charges: their exchange interaction with nucleus of atoms of matter
in a gas phase (or in residual gas of technical vacuum) is the reason for
`isotope anomaly' and `$\lambda _{\text{T}}$-anomaly' of the
orthopositronium (see [10-12]). It is possible to see idea of long-range for
baryon charge in the following fragment of the message of R.P. Feynman ``The
quantum theory of gravitation'' at Warsaw conference in July, 1962 [24]
(italics and underlining ours): `{\it There is theory, more well-known to
meson physicists, called the Yang-Mills theory, and I take the one with zero
mass; it is a special theory that has never been investigated in great
detail. It is very analogous to gravitation; instead of the coordinate
transformation group being the source of everything, it's the isotopic spin
rotation group that's the source of everything. It is a non-linear theory,
that's like the gravitation theory, and so forth. At the suggestion of
Gell-Mann I looked at the theory of Yang-Mills with zero mass, which has a
kind of gauge group and everything the same; and found exactly the same
difficulty. And therefore in meson theory it was not strictly unknown
difficulty, because it should have been noticed by meson physicists who had
been fooling around the Yang-Mills theory. They had not noticed it because
they're practical, and the Yang-Mills theory with zero mass obviously does
not exist, because a zero mass field would be obvious; \underline{it would
come out of nuclei right away}. So they didn}'{\it t take the case of zero
mass and investigate it carefully}'.

It is possible to think, that the mass-less field with a property of the
long-range for baryon charge was really showed in the orthopositronium
anomalies [10-12].

Polarization of the ``ingredients'' of its mass (M%
\mbox{$>$}
0 and M%
\mbox{$<$}
0) in ground laboratory gravitation field in conditions of a counteraction
of an external electric field [1] and the further substantiation of new
physics by old and later results of the fundamental theory will be
considered in the following section (III).

So, on an experimental basis (``$\lambda _{\text{T}}$-anomaly'' and
``isotope anomaly'' {\it o}-Ps [10-12]) it is postulated here the compound
nature and ambiguity of the fundamental space-like structure in the final
state of $\beta ^{+}$-decay of nuclei-sources of positrons and equality of
each of making mass on absolute value to the Planck mass
$$
\text{M}_\mu =\pm \text{M}_{\text{Pl}}=\pm \sqrt{\hbar c/G},
$$

\noindent so owing to infringement of a full relativity and scale invariance
at length $\sim $~2R$_\mu $ its resulting mass (``zero-mass Goldstone
field'' [20]) -- not zero, but extremely small size -- m$_\mu =\hbar /$2R$%
_\mu \cdot c\sim 2\cdot 10^{-10}$eV (see lower -- definition of
macroscopical space-like structure radius R$_\mu $).

In known discussion of a maximon problem [25] only positive value Planck
mass was taken into account, though the negative mass is formally also
possible.

In QED the reality of {\it o}-Ps superfine structure level shift is
established (`new force of annihilation' [26])
$$
(3/7)\cdot \triangle \text{W=}(3/7)\cdot \text{W}\alpha ^2\cong 3.6\cdot
10^{-4}\text{eV,}
$$

\noindent where W$\cong 6.8$ eV is a positronium binding energy; the
attraction of an electron and a positron charges in {\it o}-Ps is weakened,
as during time $\triangle t_{\text{v}}\sim \hbar /(3/7)\cdot \triangle $W
the truly neutral triplet quantum system $^3$(e$^{+}$e$^{-}$)$_1$ exists in
the form of one virtual photon, i.e. in `state' without electric charges. In
view of the additional version of the `mirror Universe' (without breaking of
{\it CP}-invariance, see [11,12]) and $^{\text{T}}$Ps$\Leftrightarrow ^{%
\text{T}}$Ps$^{\prime }$($^{\text{S}}$Ps$^{\prime }$) oscillations (the
stroke means an a belonging to the `mirror Universe'), this fact can be
interpreted as the impossibility to locate the center of $^{\text{T}}$Ps$%
\Leftrightarrow ^{\text{T}}$Ps$^{\prime }$($^{\text{S}}$Ps$^{\prime }$) in
space of the real observer within the limits of volume, smaller $\Delta ^3$;$%
\,\,\Delta $ -- virtual fundamental length (`shift')
$$
\Delta \sim c\cdot \triangle t_{\text{v}}=\frac 4{\alpha ^4}\left( \frac
\hbar {m_ec}\right) \cong \text{5.5}\cdot \text{10}^{-2}\text{cm. }\eqno (1)
$$

In standard QED negative mass relate to ``pathological'' states, which
according to the taken roots representations are not realized physically,
otherwise such physical state would be unstable in relation to a spontaneous
birth of the big number of particles (disintegration of vacuum) [27]. It is
shown below as, using (1) to postulate a natural boundary condition which
limits disintegration of vacuum: there is a reorganization of vacuum in
final state of $\beta ^{+}$-decay in the limited `volume' of space-time
(TQT). The substantiation assumed regularization, i.e. exception of the
disintegration of the vacuum with a preservation of the direct
interpretation of the negative sizes of energy and action in the ``mirror
Universe'' can be received with attraction of M. Born's ``principle of
reciprocity'' [28]:

`{\it General relativity is concerned only with point transformations in the
x-space
\mbox{$<$}
...%
\mbox{$>$}
It has struck me that the point transformations in the p-space might be
considered in the same way. Thus one is led to a kind of inverted relativity
formalism in the p-space in which everywhere space-time and momentum-energy
are interchanged. The principle laws of quantum mechanics, such as the
commutation laws, the relations of indeterminacy, etc., are symmetrical in x}%
$_k$ {\it and} {\it p}$_k$.

{\it These facts suggest very strongly the formulation of a `principle of
reciprocity', according to which each general law in the x-space has an
`inverse image' in the p-space, in the first instance the laws of
relativity'.}

Supersymmetric degeneration of the ortho- and parasuperpositronium ($%
\triangle $W$\cong 0$, see [11,12]) can be realized at enough big n=N
$$
\text{W}_{\text{N}}=\frac{e^4m_e}{4\hbar ^2\text{N}^2}\cong 0\eqno (2),
$$

\noindent where W$_{\text{N}}$ is the binding energy of N-th positronium
state. Expansion of a principle of reciprocity allows to formulate a natural
boundary condition for the completely degeneracy Fermi-gas with boundary
energy $\varepsilon _F$ (Fermi's level) [29] in the discrete {\it x}-space
$$
\varepsilon _F=(3\pi ^2)^{2/3}\cdot \frac{\hbar ^2}{2m_e}\left( \frac{\text{N%
}^{(3)}\,}{\text{V}}\right) ^{2/3}=(3\pi ^2)^{2/3}\cdot \frac{\hbar ^2}{2m_e}%
\frac 1{\Delta ^2}
$$

\noindent in view
$$
\varepsilon _F=\text{W}_{\text{N}}\text{,}\eqno (3)
$$

\noindent as N$^{(3)}$ is number of cells in the {\it p}-space, displayed in
the {\it x}-space in a volume V of the fundamental space-like structure. The
condition (3) unifies standard quantization of the atom states and
quantization of the {\it x}-space postulated here. This postulate a
transition from linear sequence of the main quantum number in the atom (n =
1, 2, 3,..., N) to number of the cells (``units'') of the 3-dimensional
space-like structure (``atom of long-range'') N$^3$ -- is a designated in
formulas N$^{(3)}$. From (2) and (3) we receive values:

\noindent -- 3-dimensional fundamental space-like structure cells number
$$
\text{N\thinspace }^{(3)}=\frac{2^{9/2}}{3\pi ^2\cdot \alpha ^9}\cong
1.302\cdot 10^{19}\eqno (4)
$$

\noindent -- the linear area 2R$_\mu $ of the fundamental space-like
structure with the center in ``point'' of $\beta ^{+}$-decay during time
where R$_\mu $ is the Bohr radius of N-th positronium state
$$
\text{r}_{\text{N}}=\frac{2\hbar ^2\text{N}^2}{e^2m_e}\cong 5.57\cdot 10^4%
\text{cm}\equiv \text{R}_\mu
$$

and
$$
\tau _\mu =\frac{\text{R}_\mu }c\cong 2\cdot 10^{-6}\text{s.}\eqno (5)
$$

If we ``occupy'' each cell with a natural structural unit of the stable
matter ``electron(e)/proton(p)'' for M$_\mu $
\mbox{$>$}
0 and ``electron hole ($\overline{e}$)/proton hole ($\overline{p}$)'' for M$%
_\mu $
\mbox{$<$}
0, then we shall receive the fundamental mass
$$
\begin{array}{c}
\text{M}_\mu =\text{N\thinspace }^{(3)}\cdot (m_p+m_e)=\frac{2^{9/2}}{3\pi
^2\cdot \alpha ^9}\cdot (m_p+m_e)\cong \\ \cong 2.179\cdot 10^{-5}\text{g.}
\end{array}
\eqno (6)
$$

\noindent Comparison of the received value M$_\mu $ with Planck mass
obviously
$$
\pm \text{M}_{\text{Pl}}=\pm \sqrt{\hbar c/G}\cong 2.177\cdot 10^{-5}\text{g.%
}
$$

At a conclusion (6) canonical values of positronium binding energy and the
size are accepted (Bohr's atom, Schrodinger equation). The account of the
radiation corrections improves the accuracy of the representation of Planck
mass (and gravitational constant {\it G}) through a constant of thin
structure $\alpha $.

Thus, identifying the received value of two-digit fundamental mass M$_\mu $
with Planck mass (with accuracy $\sim $~0.1\%) on the basis of experiment
[10-12] and additional realization of supersymmetry, when observable there
is a shift (non-locality), and superpartners are latent from supervision in
the `mirror Universe', we are postulate `{\it additional G}$\hbar /c$-{\it %
physics}'.

\section{ON THE SUPPRESSION MACROSCOPIC QUANTUM EFFECT BY ELECTRIC FIELD}

\medskip Since second half of the 1970th the experimental program has been
proved which result became an establishment of ``isotope anomaly'' lifetime
spectra of the positrons annihilation from $\beta ^{+}$-decay $^{22}$Na in a
gaseous neon ($\gamma _{\text{n}}$-$\gamma _{\text{a}}$-delayed coincidence;
$\gamma _{\text{n}}$-- is the nuclear gamma-quant, $\gamma _{\text{a}}$--
one of annihilation gamma-quanta) [10]. A growth of the intensity
long-living orthopositronium components of annihilation lifetime spectra was
observed (the factor 1.85$\pm $0.1) during a reduction of the contents of an
isotope $^{22}$Ne from 8.86\% (in neon of natural isotope composition) up to
4.91\%. The standard estimation of the effect from isotopic shift of a
threshold of the positronium formation in these conditions gives a
vanishingly small value $\sim $10$^{-6}$.

Full degeneration of the ground states para- and ortho-superpositronium in
{\it N}=2 supersymmetric QED ({\it N}=2 SQED) [30], as a manifestation of
{\it o}-Ps oscillation in the `mirror Universe' (with conservation of {\it CP%
}-parity, see [11, 12]), it is based phenomenology of the {\it o}-Ps
`isotope anomaly', and old calculations {\it o}-Ps annihilation on a
single-quantum and the neutral spin 1 boson U supersymmetric theory [17]
have allowed to receive a realistic estimation of the contribution of an
additional annihilation mode $^{\text{T}}$Ps$\backslash ^{\text{T}}$Ps$%
^{\prime }$($^{\text{S}}$Ps$^{\prime }$) $\sim 2\cdot 10^{-3}$ (``$\lambda _{%
\text{T}}$-anomaly''). This result can be interpreted as a consequence of
the topological quantum transition in $\beta ^{+}$-decay (see [11, 12]).
Thus, in works [11, 12] the substantiation and the quantitative description
of the observable macroscopical quantum effects in final state of $\beta
^{+} $-decay of the nuclei such as $^{22}$Na, $^{68}$Ga ($\triangle $J$^\pi $
= 1$^{+}$) is received. In the previous part (II) the conceptual foundations
examined statement of a question are stated.

In article [31] (see section I) the critical analysis of the conclusions of
the work [1] is carried out. It has been shown, that a technique of the new
experiment of Michigan group, in which the electric field is used, is
qualitatively distinct from the techniques of the works [5, 6]. A
theoretical base of the critical analysis [31] (in section I) of Michigan
group conclusions [1] from the positions of the quantum theory of field
(QTF) is:

$\bullet $ hypothesis about additional realization of a supersymmetry
(``additional $G\hbar /c$-physics'');

$\bullet $ antipodal symmetry of an energy (mass) and action in the ``mirror
Universe'' in relation to the real observer (A.D. Linde [32]). For a
consideration of vacuum reconstruction in the limited ``volume'' of
space-time in $\beta ^{+}$-decay the final state from the positions of the
GR concepts are incorporated (see section II):

$\bullet $ spontaneously broken full relativity (A.F. Andreev, 1982);

$\bullet $ vacuum-like state of matter (VSM, E.B. Gliner, 1965) and

$\bullet $ discrete, scalar C-field with the negative density of the energy
(mass), compensating VSM (F. Hoyle \& J.V. Narlikar, 1964).

For an explanation of the orthopositronium anomalies [5,6,10] the model of a
long-range for a baryon charge is offered, as `...a qualitatively allocated
state of the environment is the special phase, basically capable to exchange
energy, pulse, baryon charge, etc. with the other phases of the environment'
[33]. The limit sized VSM (we shall name it ``an atom of the long-range'')
has the discrete, crystal-like structure with the common number of the cubic
cells (``units'') M$_{\text{Pl}}$ /($m_P$ $+$ $m_e$)$\,\cong $ $1.302\cdot
10^{19}$ (M$_{\text{Pl}}$ is Planck mass, $m_p$ -- proton mass, $m_e$ --
electron mass). In an initial state (trivial topology) all types of the
charges in ``units'' VSM are compensated by the charges of an opposite sign
on the C-field the (the ``mirror Universe''). In final state of the $\beta
^{+}$-decay in a gravitational field in the {\it o}-Ps self-annihilation
time occurs decompensation of a baryon charge, as the VSM (a positive mass M$%
_{\text{Pl}}$) falls in a gravitational field, and the ``mirror Universe''
with the negative mass --$\left| \text{M}_{\text{Pl}}\right| $ goes
opposite. A conformity of the structures of an usual substance
(``atom''/``nucleus of atom'') and VSM takes place: in ``an atom of the
long-range'' (2R$_\mu \sim 1.2\,$km) are allocated a ``nucleus of atom of
long-range'', containing $\overline{\text{n}}=5.2780\cdot 10^4$%
cells-``units'' (2{\it r}$_{\overline{\text{n}}}\sim 2.6\,$cm; {\it r}$_{%
\overline{\text{n}}}\,$-- radius of ``nucleus''), with wich {\it o}-Ps
interact [11,12].

In the ``additional $G\hbar /c$-physics'' observable there is a space
``shift'' in the constitution of the VSM's discrete structure (supersymmetry
is restored). In a Standard Model supersymmetry is broken, space shift is
not examined as an observable and it is supposed, that the mechanism of the
supersymmetry breaking will be established as a result of the supervision of
superpartners.

Sequence of the ``shifts'' in a process of the oscillations $^{\text{T}}$Ps$%
\backslash ^{\text{T}}$Ps$^{\prime }$($^{\text{S}}$Ps$^{\prime }$) forms a
Hamiltonian cycle (in the language of the graph theory), connecting a
``nucleus of an atom of the long-range'' and lifetime of {\it o}-Ps:
``wandering'' {\it o}-Ps at its oscillations to the `mirror Universe' till
the annihilation moment can be presented as a way with an steps length on
all ``units'' (without recurrences) in a ``nucleus of an atom of the
long-range''. It assumes calculation of the shares ($l_{\text{a}}$, $l_{%
\text{b}}$, $l_{\text{c}}$) each of the possible steps in total length of
the way L
$$
\text{L =\thinspace }\overline{\text{n}}\cdot \Delta \cdot \left( \frac{l_{%
\text{a}}}{l_{\text{a}}+l_{\text{b}}+l_{\text{c}}}+\frac{l_{\text{b}}}{l_{%
\text{a}}+l_{\text{b}}+l_{\text{c}}}+\frac{l_{\text{c}}}{l_{\text{a}}+l_{%
\text{b}}+l_{\text{c}}}\right) .
$$

In a spatial cubic lattice, alongside with steps on the edges of a cube (a =
1) still two type steps are possible: on a small diagonal (on the verge of a
cube) b =$\sqrt{2}$ and on the big diagonal (between as much as possible
removed tops of a cube) c =$\sqrt{3}$. It is equivalent to appearing of two
allocated speeds $c$/$\sqrt{2}$ and $c$/$\sqrt{3}$ along with the light
velocity. It is interesting, that it is recently marked peculiarity of
speeds $c$/$\sqrt{2}$ and $c$/$\sqrt{3}$ in GR [34]. It is possible to make
the following estimations for minimal (a) and maximal (c) distances between
units. For ``a nucleus of an atom of the long-range'' the prospective length
Hamiltonian cycle gives an estimation of the orthopositronium lifetime:

\noindent from below
$$
\frac{\overline{\text{n}}\cdot \Delta \cdot \text{a}}c\cong \frac{52780\cdot
5.5\cdot 10^{-2}\text{cm}}{3\cdot 10^{10}\text{cm/s}}\cong 96.9\cdot 10^{-9}%
\text{s}
$$

\noindent and from above
$$
\frac{\overline{\text{n}}\cdot \Delta \cdot \text{c}}c\cong \frac{52780\cdot
5.5\cdot 10^{-2}\cdot \sqrt{3}\text{cm}}{3\cdot 10^{10}\text{cm/s}}\cong
167.8\cdot 10^{-9}\text{s}
$$

$$
96.9\,\text{ns
\mbox{$<$}
\mbox{$<$}1/}\lambda _{\text{T }}(\cong 142\,\text{ns)
\mbox{$<$}
\mbox{$<$}}167.8\,\text{ns.}
$$

These estimations are the additional argument of the allocation of ``a
nucleus of an atom of the long-range'' in structure ``an atom of the
long-range''.

The description of an additional mode ($\sim $ 0.2\%) {\it o}-Ps
annihilation [12] means realization, along with standard QED -- one of the
fundamental theories in a structure of the ``cube of the physical theories''
(M.P.Bronshtein, 1934/A.L. Zelmanov, 1967), describing dynamics of 99.8\%
{\it o}-Ps, -- and a new physdics. The new physics (``additional -physics'')
corresponds to ``top'' (0,$\hbar ,G$) of the ``cube of the physical
theories'', representing `...non-relativistic quantum gravitation (NQG),
\mbox{$<$}
...%
\mbox{$>$}
concerning which it is not clear, whether there are objects, which it
describes...' [35]. By A.L. Zelmanov identified all tops of ``cube'', except
of this.

The analysis with the account for ``an additional $G\hbar /c$-physics''
shows, that the decision of the orthopositronium problem can be received as
a result of the comparative measurements which are not demanding changes of
a technique of work [1]. It is necessary to compare with the other things
being equal, the results of two measurements, when:

\noindent $\bullet $ the vector of an intensity of an electric field ${\bf E}
$ is directed parallel to a gravity ${\bf P}$ (the perpendicularly to
flatness of a porous film in which it is formed Ps; though in the published
work of Michigan group are not present the information of the mutual
orientation of the vectors ${\bf E}$ and ${\bf P}$, but the standards of an
illustration of a scientific texts allow to assume according to Fig. 1 of
[1], that this variant of experiment is already realized, in which the
additional supersymmetric mode with the contribution $\sim $~$2\cdot 10^{-3}$
is suppressed) and

\noindent $\bullet $ the vector of an intensity of an electric field ${\bf E}
$ is directed perpendicularly to a gravity ${\bf P}$ and to flatness of a
porous film. Let's calculate intensity E of the electric field necessary for
a deviation of a beam of positrons with energy 700 eV on a porous film. In
non-relativistic approximation the positron trajectory is determined by
expression (see, for example, [36])
$$
\text{x}=\frac{m_e\cdot \left| e\right| \cdot \text{E}}{2\text{p}_{\text{y}%
}^2}\cdot \text{y}^2+\frac{\text{p}_{\text{x}}}{\text{p}_{\text{y}}}\cdot
\text{y .}
$$

Let's choose an axis x$\parallel $E (projection of a positron pulse p$_{%
\text{x}}$=0). Then in view of geometry of the measuring cavity used in work
[1] we receive an intensity of an electric field E$_{\text{exp}}$
\mbox{$<$}
4$\cdot $10$^3$V/cm.

Opposite charged matrixes -- cubic lattices of ``particles'' in a structure
of a macroscopical quantum structure VSM and ``holes'' in a structure
``mirror Universe'' (accordingly, positive and negative mass) -- are
accelerated by an electric field in one direction. Action of an electric
field is shown in ``units'' which represent ``points'' of a crossing of the
observable Universe (VSM) and C-fields (``mirror Universe''). Gravitation,
on the contrary, moved apart these matrixes in a vertical direction. This
counteraction is shown on a Fig.: electric force (2e${\bf E}$ on ``unit'')
operates on an electric dipole under a angle (180$^{\text{o}}$ -- $\alpha $
) in relation to a gravity ${\bf P}$, counteracting decompensation of a
baryon charge.

\begin{picture}(100,120)
\put(200,85){\vector(1,1){20}}
\put(200,85){\vector(0,-2){40}}
\end{picture}
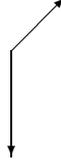
Fig. Direction of electric force relation to gravity

\bigskip In order to opposite sign baryon charges (and connected with these
electric charges) did not compensate each other in each ``unit'', they
should moved apart (on a vertical) on the distance equal to the baryon size
(``proton'' in a structure VSM). The characteristic size of a proton is
determined by its Compton's length
$$
\lambda _{\text{p}}=\frac \hbar {m_p\cdot c}\approx 2\cdot 10^{-14}\text{cm.}
$$
\noindent Hense, for the decompensation at would be an actions of an
effective (critical) acceleration of the mass $\gamma _{\text{cr}}$ during
lifetime {\it o}-Ps ($\tau _{\text{T}}$ $\cong 1.42\cdot 10^{-7}$s), enough
$$
\frac{2\gamma _{\text{cr}}\cdot \tau _{\text{T}}^2}2\cong 2\cdot 10^{-14}%
\text{cm, }\gamma _{\text{cr}}=\frac{2\cdot 10^{-14}}{\tau _{\text{T}}^2}%
\cong 1\text{cm/s}^2.
$$

\noindent Other estimation $\gamma _{\text{cr}}$, reflecting an exchange
interaction of a baryon charges of nucleus of atoms of a matter with a
baryon charge in ``unit'', accounts the radius of an action of a nuclear
forces r$_{\text{b}}=(2\div 3)\cdot 10^{-13}$cm
$$
\gamma _{\text{cr}}=\frac{(2\div 3)\cdot 10^{-13}}{\tau _{\text{T}}^2}\cong
10\,\text{cm/s}^2.
$$

The distinction between the resulting estimations on the order on an end
result will not be reflected, because the acceleration of a gravity g
\mbox{$>$}
\mbox{$>$}
$\gamma _{\text{cr}}$. It is necessary to take into account, that there are
under an action of a gravitational field all ``atom of the long-range'' (M$_{%
\text{Pl}}$), i.e. $\sim $~10$^{19}$ of its ``units'', and the external
electric field operates only in a volume of the measuring chamber V$_{\text{%
exp}}$ = (1.5 cm)$^3\cong $ 3.4 cm$^3$ [1]. In this volume getting in a
sphere of a ``nucleus of an atom of the long-range'', contains n = V$_{\text{%
exp}}$/$\Delta ^3=$ 3.4 cm$^3$/($5.5\cdot 10^{-2}$cm)$^3\cong 1.7\cdot 10^4$
``units''. Let's notice, that all volume of the measuring chamber is
included to the structure of a ``nucleus of an atom of the long-range'', as $%
\overline{\text{n}}\,$($5.2780\cdot 10^4$)
\mbox{$>$}
n($1.7\cdot 10^4$). Therefore, operating with the force, braking falling
VSM, an electric field creates the braking equal
$$
\frac{2\text{n}\cdot e\text{E}\cdot \cos \alpha }{m_{\text{eff}}}\,,
$$
\noindent where $m_{\text{eff}}$ is an effective mass.

There is a question how to determine m$_{\text{eff}}$? As at the ``top'' NQG
are simultaneously realized both ``classics'' and ``quanta'' then the mass
applies for the role of this size is only
$$
m_{\text{S}}=e/\sqrt{G}\sim 10^{-6}\text{g,}
$$

\noindent for the first time considered by G.J. Stoney (1881) long before
opening a constant of Planck (see [37]). An expression for $m_{\text{S}}$
has been received by an equating of the potentials Coulomb and Newton, a in
spirit of a counteraction of an electric force and gravitation examined
here:
$$
\frac{e^2}{\text{r}}=G\cdot \frac{m_{\text{S}}^2}r\,.
$$
Let's, in passing notice, that the squares of the charges -- an electric and
``gravitational'' -- mean, as well as for Planck mass, an opportunity of
their treatment as a plus- and minus-particles. Now there is all for the
quantitative formulation of a condition of a suppression by an electric
field of the additional {\it o}-Ps self-annihilation mode:
$$
\text{M}_{\text{Pl}}\left( g-\frac{2\text{n}\cdot e\text{E}\cdot \cos \alpha
}{m_{\text{eff}}}\right) <\text{M}_{\text{Pl}}\cdot \gamma _{\text{cr}}
$$
\noindent also we receive criterion of the suppression
$$
\text{E
\mbox{$>$}
\mbox{$>$}}\frac{(g-\gamma _{\text{cr}})\cdot m_{\text{eff}}}{2\text{n}\cdot
e\cdot \cos \alpha }=\frac{g-\gamma _{\text{cr}}}{2\overline{\text{n}}\cdot
\sqrt{G}\cdot \cos \alpha }\,,
$$
\noindent as $m_{\text{eff}}\equiv m_{\text{S}}$, n $\rightarrow \overline{%
\text{n}}\,$.\thinspace \thinspace At $\alpha \cong 0\,\,$({\bf E} in
anti-parallel {\bf P}) the intensity of an external electric field should
exceed critical value E$_{\text{cr}}$ $\sim $~10$^4$ V/cm. The intensity of
an electric field in the experiment [1] E$_{\text{exp}}$ satisfies to the
criterion of the suppression of an additional supersymmetric mode of the
{\it o}-Ps annihilation E$_{\text{exp}}\sim \,\,$E$_{\text{cr}}$\thinspace
\thinspace .

In the conditions when {\bf E} is perpendicular to {\bf P} ($\alpha \cong
90^{\text{o}}$, cos $\alpha \cong $0, E$_{\text{exp}}$%
\mbox{$<$}
\mbox{$<$}
E$_{\text{cr}}$), the additional supersymmetric mode with the contribution $%
\sim $~$2\cdot 10^{-3}$ should be restored.\smallskip

\noindent {\it \underline{Addition at a proof-reading}\medskip}

The opportunity to check up the assumption of a conditions of a measurements
in work [1] (see above) under the dissertation of R.S. Vallery ``RESOLUTION
OF THE ORTHOPOSITRONIUM LIFETIME PUZZLE'' (University of Michigan, 2004) has
appeared. A comparison of the diagram (Fig.1 in [1]) with photos of the
installation and the measuring chamber (Fig. 4.6 and Fig. 4.15 in the
dissertation) leaves no doubt that the measurements were carried out in
conditions, when an electric field ${\bf E}$ is in a parallel to gravity $%
{\bf P}$.

Any of the methods and tests for a substantiation and exceptions of the
regular mistakes submitted in the dissertation in details, does not
contradict alternative interpretation of the result, received in a vacuum
experiment with an electric field [1] discussed here. Additional
annihilation mode with the contribution ($0.19\pm 0.02$)\% in the
experiments in a magnetic field with buffer gases [5] and with the
contribution ($0.14\pm 0.023$)\% in vacuum experiment without an electric
field [6], it is caused by non-local interaction {\it o}-Ps with VSM and the
C-field (the ``mirror Universe''). In the new vacuum experiment the
additional mode is suppressed by action of an electric field to the
macroscopical quantum structure ``VSM + C-field''. It can be restored by
``neutralization'' of an action of an electric field at $\alpha \cong 90^{%
\text{0}}$\thinspace \thinspace the all intensities E$\sim 10^4$V/cm. This
alternative interpretation of the all results of Michigan group, received
for two decades, cannot exclude even measurement of a kinetic energy {\it o}%
-Ps according to a time of the flight, because the small share {\it o}-Ps ($%
\leq 0.2\%$) represents a component (the ``phase'') cooperating with a
macroscopical quantum state (``VSM + C-field'') in a final state of the $%
\beta ^{+}$-decay and dropping out of the standard kinetics laws (see
section I). The situation with {\it o}-Ps is in this respect similar to
two-component structure of a superfluid state of the liquid helium. The
tests, used in the dissertation, cannot allocate the small, but very
important contribution from the basic point of view of this ``phase'' of
{\it o}-Ps. The choice for the benefit of one of alternatives without basic
change of a technique can be made only on the basis of comparative
measurements of the {\it o}-Ps self-annihilation decay rate at ${\bf E}$
parallel and ${\bf E}$ perpendicular to gravity ${\bf P}$ under other
identical conditions.

\medskip We thank S.G. Karshenboim for the given access to R.S.Valleri's
dissertation.$\bigskip$

$\dagger \,\,$Year ago has died Dr. Sci. (Phys.\&Math.) Boris Aleksandrovich
Kotov (19$\frac 8{\text{X}}$38 -- 20$\frac{10}{\text{IV}}$05). This work and
the forthcoming experimental check of the ideas, which are put in its basis,
all this -- last scientific problem which interested B.A. till final days.
The basic feature of B.A.Kotov's person -- Physicist-Constructor-Manager --
was the feeling new. But novelty should be well proved by previous
experience of a science and practice. From the end of 1980th, being already
the Head of the Research-and-Production Association ``Electron Integral
Systems'' (Saint-Petersburg), B.A. has constantly supported these researches.

This article has been published year ago in a format of the Preprint \#1784
A.F.Ioffe Physical Technical Institute of Russian Academy of Science.
\newline

\noindent
$^{1}$ E-mail address: bormikhlev@mail.ioffe.ru \newline
$^{2}$ E-mail address: v.sokolov@mail.ioffe.ru

\end{document}